# Complex Clipping for Improved Generalization in Machine Learning

Les Atlas, *Life Fellow, IEEE*, Nicholas Rasmussen*, Student Member, IEEE,* and Felix Schwock, *Student Member, IEEE*, Mert Pilanci, *Member, IEEE*

*Abstract*—For many machine learning applications, a common input representation is a spectrogram. The underlying representation for a spectrogram is a short time Fourier transform (STFT) which gives complex values. The spectrogram uses the magnitude of these complex values, a commonly used detector. Modern machine learning systems are commonly overparameterized, where possible ill-conditioning problems are ameliorated by regularization. The common use of rectified linear unit (ReLU) activation functions between layers of a deep net has been shown to help this regularization, improving system performance. We extend this idea of ReLU activation to detection for the complex STFT, providing a simple-to-compute modified and regularized spectrogram, which potentially results in better behaved training. We then confirmed the benefit of this approach on a noisy acoustic data set used for a real-world application. Generalization performance improved substantially. This approach might benefit other applications which use time-frequency mappings, for acoustic, audio, and other applications.

## I. INTRODUCTION

MANY modern systems for classification and estimation make use of deep networks. These trainable systems are typically overparameterized, with many more network parameters (weights and biases) than training examples. As has been discussed by others, due to this overparameterization, deep networks have many "critical points," which can, for example, be local minima [1] or saddle points [2]. During training, these critical points are singularities in neural networks that often detract from learning, as singularities do for other ill-posed training examples. As such, singular behavior within neural networks has been a rich topic of recent research. For example, methods to regularize and reduce their prevalence [3], methods to avoid singularities within the network [4], and the effect of these singularities on the dynamics of learning within multilayer perceptrons [5]. Most of this work argues that singularities should be avoided or ameliorated by reducing the possibility of linear dependencies within and network layers. These linear dependencies can cause learning difficulties, such as non-smooth loss functions, where minimizers are not well-behaved; hence generalization becomes more difficult [6].

## II. CONTRIBUTION

In this letter, we consider and reduce linear dependencies in a different part of the network, the input representation. At first consideration, this reduction is a very difficult problem since most real-world signals are flush with natural sources of linear dependencies. For example, acoustic propagation is subject to natural reverberation, and radio communication signals are subject to multipath, both of which introduce linear dependencies. Most pre-processing, such as changing sample rate and filtering, will also contribute to linear dependencies. Even the short-time Fourier transform, with its usual overlap/add processing, has implicit linear dependencies between frequency channels. This letter does not model those linear dependencies. We instead consider reducing linear dependencies by carefully considering and modifying the usual input processing of acoustic and similar signals.

Many deep nets systems have inputs from Fourier transforms, such as spectrograms, mel-frequency spectral coefficients, and other frequency representations of real sequences. These representations usually use real numbers that arise from usual magnitude operations on initial short-time Fourier transforms, which are typically complex-valued. This conversion from complex values to real numbers, and possible small extra processing to remove linear dependencies and hence singularities, is the focus of this letter. In particular, we will suggest how modifications to standard magnitude detection, as used, for example, in converting the complex output of a short-time Fourier transform to a non-negative real spectrogram, can remove much of the linear dependence in training and test data.

This project has been funded in part with Federal funds from the Department of Health and Human Services; Administration for Strategic Preparedness and Response; Biomedical Advanced Research and Development Authority, under Contract No. 75A50122C00034.

Les Atlas is with the Department of Electrical and Computer Engineering, University of Washington, Seattle, WA, 98105 (e-mail: atlas@uw.edu).

Nicholas Rasmussen is with The Covid Detection Foundation DBA Virufy, Los Altos, CA and The University of South Dakota, Department of Computer Science, 414 East Clark Street, Vermillion, SD (e-mail: nicholas.rasmussen@virufy.org).

Felix Schwock is with the Department of Electrical and Computer Engineering, University of Washington, Seattle, WA, 98105 USA (e-mail: fschwock@uw.edu).

Mert Pilanci is with the Department of Electrical Engineering, Stanford University, Stanford, CA 94305 USA (e-mail: pilanci@stanford.edu).

The main contribution of this letter is to show how to remove this deleterious linear dependence in spectrograms and similar deep net input representations–while retaining all remaining independent information encoded by the spectrogram. Furthermore, this new form of regularization produces improved conditioning–that is shown via improved condition numbers of the input arrays, and substantially improved generalization performance on a real-world example of a noisy and important data application.

### III. THE SHORT TIME FOURIER TRANSFORM (STFT)

The conventional approach to classification of acoustic and similar signals with deep nets has predominantly used time-frequency representations such as spectrograms. Spectrograms typically are found via a detection operation on a short time Fourier transforms, often defined with a mixed discrete time ($n\epsilon\mathbb{Z}$) and frequency ($k\epsilon\mathbb{Z}$) notation as

$$X[n,k] = \sum_{m=0}^{K-1} w[m]x[m+nH]e^{-j\frac{2\pi}{K}km}.$$

There are $L$ input points $x[n]_{n=0,\ldots,L-1} = \mathbf{x} \in \mathbb{R}^L$ available for each overall STFT calculation, which produces a matrix of complex values of size $(K/2) \times (M+1)$, where only non-negative frequency indices were used. The complete STFT input uses a sequence of $M$ shifting steps, called "frames," which shift over $\mathbf{x}$. $H$ is the amount of shift, often called "hop size." The data window is optionally weighted by a tapered window $w[n]$ of length $K$. (Mel-spaced spectrograms and Mel frequency cepstral coefficients modify this representation by using non-uniform spacing in $k$.) This STFT uses the above Fourier sum to map typically real one-dimensional values to complex 2-dimensional values $\mathbb{R}^L \to \mathbb{C}^{(K/2)\times(M+1)}$.

Typically, $(K/2) \times (M+1)$ is greater than or much greater than the size of the input $\mathbf{x}$, which is $L$ sequential points. Thus, the STFT's real to complex mapping is typically heavily redundant. While the STFT is usually a beneficial step for machine learning since it breaks the input into frequency channels, this benefit potentially comes with the cost of added redundancy. It would thus be expected that deleterious linear dependence is already within the input $\mathbf{x}$ and is also potentially increased by the STFT. It is also clear that these processing parameters can influence the linear dependence between values of $X[n,k]$, which suggests that operations on $X[n,k]$ are a particularly good place to remove this deleterious dependence.

### IV. PROBLEM STATEMENT: A JOINT REGULARIZER AND DETECTOR

The STFT is often used as a precursor to a spectrogram. To make use of existing successful deep net architectures, a STFT needs a detection operation to form the non-negative real values of a spectrogram $\mathbb{C}^{(K/2)\times(M+1)} \to \mathbb{R}_+^{(K/2)\times(M+1)}$, where $\mathbb{R}_+$ are non-negative real numbers, via a magnitude square $|X[n,k]|^2$, usually followed by a scaled logarithm, providing the common 2-dimensional spectrogram in decibels. Without loss of generality, this scaling factor can equate magnitude square and magnitude (rectification) detection.

In this letter, our main deviation from standard practice is to generalize this magnitude square detection operation in such a way that it removes linear dependence. As with the usual magnitude square for the spectrogram, the detection operation maps a complex range to a continuous non-negative real range $f(X[n,k])$, $\mathbb{C} \to \mathbb{R}_+$.

This STFT detection has an interesting similarity with activation functions in deep nets. The most common activation function, ReLU (rectified linear unit) activation function is defined as $g(x) = \max(0,x)$, $\mathbb{R} \to \mathbb{R}_+$ where $x$ is the typically real input to the activation function. (Complex inputs will be considered starting in Section D below.) Numerous studies with deep nets have shown how these ReLU activation functions have advantages over previous smooth activation functions such as sigmoid shapes. It should also be noted that, like a magnitude or magnitude-squared detector of a spectrogram, a ReLU maps to non-negative real numbers. Some of the benefits of the ReLU function over other deep net activation functions are:

#### A. Better performance, especially when combined with dropout

There are numerous examples where ReLU activations performed better than other activation functions. For example, reference [7] trained rectifier networks with up to 12 hidden layers on a proprietary voice search dataset containing hundreds of hours of training data. After supervised training, rectifier deep nets performed substantially better than their sigmoidal counterparts. [8] applied deep nets with ReLU activations and dropout regularization to a segment of broadcast news that is a large vocabulary conversational speech recognition task with 50 hours of training data. Rectifier DNNs with dropout, which effectively sets some internal values to zero, outperformed sigmoidal networks without dropout. The authors of this study suggested that there was: "…evidence that ReLUs and dropout have synergistic effects and we recommend that they be used together."

#### B. Reduction of linear dependence

Consistent with our overarching goals of removing all contributions to linear dependence on training and test data, in other data adaptive systems, such as acoustic echo cancellation, there is no use of deep nets. Nevertheless, these systems cancel signal paths for unknown and variable acoustic pathways [9]. In [9], the potential problem is singular behavior when trained on data, as mentioned previously [3][4][5]. This ill-conditioning issue has been considered a fundamental limitation in data-adaptive systems [10]. Yet, in [9], the authors found that introducing a nonlinearity, a half-wave rectifier, gave good performance despite the known linear dependence introduced by the variable and unknown acoustic paths. Remarkably and coincidentally, this nonlinearity is identical to the common ReLU used as activations by modern deep nets, suggesting that part of the success of ReLU activations comes from their reduction of linear dependence, be it between layers or, as we will propose, for the encoding of deep net inputs.

### C. Reduction of vanishing gradients

Sigmoidal deep nets can suffer from the vanishing gradient problem [11]. Then, like the ill-conditioning problems caused by linear dependence, the final trained network often converges to a poor local minimum. A more recent paper found that for large vocabulary speech recognition, ReLU activation functions significantly reduced word error rate as compared to sigmoidal nonlinearities [12]. [12] concluded that: "The increased sparsity and dispersion of ReLU hidden layers may help to explain their improved performance in supervised acoustic model training." This study also found that both types of rectifiers, standard ReLU, which provided zero output for negative inputs, or a leaky rectifier, which passed heavily attenuated levels for negative inputs, made no significant difference in performance. Other work has shown how fully complex-valued unitary networks can avoid the vanishing gradient problem [13] [14].

### D. Dispersion of sparse codes

It was also shown in [12] that ReLU deep nets contain substantially sparser representations than sigmoidal deep nets, producing sparse codes where information is better dispersed, and distributed more uniformly across hidden units, which is advantageous for both classifier performance and invariance to input perturbations.

We wish to generalize this usual magnitude-squared spectrogram detection operation, where we maintain the advantages of the ReLU operations, keeping the above four benefits, yet providing a mapping to interface with usual real number deep nets which already have some level of performance success with non-negative real inputs. That is, for a mapping $f(X[n,k]), \mathbb{C} \to \mathbb{R}_+$ where $X[n,k]$ represents the underlying complex STFT of the deep net input sequence and $f(X[n,k])$ replaces the usual spectrogram.

### E. Constraints on the activation function

There have been many past papers on complex activation functions for complex deep nets, $h(z), \mathbb{C} \to \mathbb{C}$. For example, [15] gives a thorough summary of possibilities. However, despite the ubiquity of spectrograms, to the best of our knowledge, magnitude or magnitude square (usually in decibels) are the only choices that have been assumed for $\mathbb{C} \to \mathbb{R}_+$ mappings from STFTs to spectrograms, with similar mappings originating with the original swept-frequency analog spectrograms which burned images into Teledeltos facsimile paper [16]. A key step in our work is to generalize and beneficially deviate from this long-held view. Desired properties for a non-negative real mapping $f(X[n,k])$ from a STFT $X[n,k]$ to a non-negative real deep net input $Y[n,k]$ are:
1) $f(X[n,k])$ is nonlinear and bounded.
2) $f(X[n,k])$ is upper/lower and right/left symmetric.
Letting $X[n,k] = Re\{X[n,k]\} + jIm\{X[n,k]\}, f(X[n,k]) = [Re\{X[n,k]\}]^2 + [Im\{X[n,k]\}]^2$. $Y[n,k] = f(X[n,k])$ is beneficially many-to-one, with $Y[n,k] = f(X[n,k]) = f(-X[n,k]) = f(X^*[n,k]) = f(-X^*[n,k])$. Viz, $f(X[n,k]), \mathbb{C} \to \mathbb{R}_+$ has even symmetry from the lower to the upper and from the left to the right half-planes of the complex STFT $X[n,k]$.

These first two constraints are the same constraints seen with the spectrogram $|X[n,k]|^2$. Yet a key change from the spectrogram is that we also wish to introduce regularization, akin to what ReLU does in one dimension, which implies the next constraint:
3) $f(X[n,k])$ has minimal non-zero support area.
Similar to the mapping of ReLU activation functions, $|Y[n,k]|^2 = f(X[n,k])$ is as sparse as possible, having a fixed but *minimal area* non-zero support region of the nonlinear mapping. This property reduces linear dependence, ideally removing only the redundant parts of the range of $f(X[n,k])$.

With the symmetries of above point 2, a triangle in the upper (and lower) half-plane would set the half of the $f(X[n,k])$ input space to zero, much as a ReLU sets half of its input space to zero. As shown in Figure 1, this constraint in two dimensions with the symmetries above, results in a cone-shaped support region where the boundary of the cone is specified by the lines in the complex plane $\{f(X[n,k]) : |Im\{X[n,k]\}| = |Re\{X[n,k]\}|\}$, depicted in Figure 1.

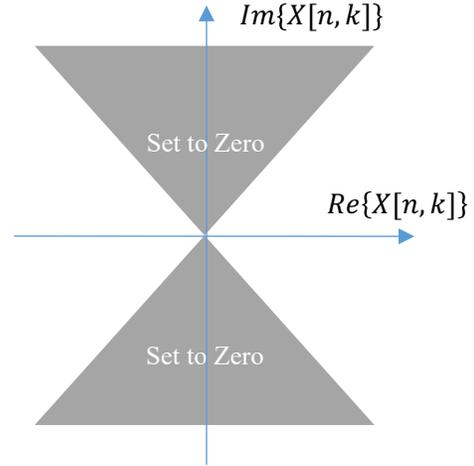

Fig. 1. Set to zero complex clipping support region of $f(X[n,k])$.

4) $f(X[n,k])$ retains zero-crossings.
The zero-crossings of a signal contain important information about its behavior and are crucial for preserving its characteristics. As demonstrated in studies such as [17] and [18], these zero-crossings can be used in the analysis and recovery of the signal. Logan's Theorem [19] provides a theoretical foundation for this recovery process. Note that the ReLU activation sets the negative half-space of its input domain, the real line, to zero. If we extend this mapping to two dimensions, it follows that the boundary of the support region $\{f(X[n,k]) : |Im\{X[n,k]\}| = |Re\{X[n,k]\}|\}$. should enclose zero values. This justifies the placement of the double cone's vertex at $|Im\{X[n,k]\}| = |Re\{X[n,k]\}| = 0$, for reasons of completeness along with reduction of redundancy.

Previous studies have demonstrated the difficulty and impracticality of recovering a signal using only its zero-crossings, as even a small amount of noise can lead to instabilities in the reconstruction (e.g., [20]). In contrast, the complex activation functions for this work retains more information than only the zero-crossings.

## V. A New Activation Function: "Complex Clipping"

With the cone's vertex at $|Im\{X[n,k]\}| = |Re\{X[n,k]\}| = 0$, and with the cone a connected region, again with analogy to ReLU, $f(0) = 0$ and for the extension to 2 dimensions, every point on the boundary of the cone should map to zero. (It is possible to consider other small modifications of this behavior, such as soft thresholding [21].) With the boundary of the cone specified by $\{f(X[n,k]) : |Im\{X[n,k]\}| = |Re\{X[n,k]\}|\}$ the equation for a proposed cone-shaped activation function is:

$$Y[n,k] = f(X[n,k]) = \begin{cases} |X[n,k]|^2, & if\ |Im\{X[n,k]\}| \leq |Re\{X[n,k]\}| \\ 0, & if\ |Im\{X[n,k]\}| > |Re\{X[n,k]\}| \end{cases} \quad (1)$$

We refer to this activation function and method of producing a regularized spectrogram as "complex clipping," as depicted in Figure 1. Since the previously discussed upper/lower and left/right symmetry implies that the region set to zero is redundant with the non-zero region above, an equivalent regularization would be an $\phi = \frac{\pi}{2}$ rotation of the cone shape implied by equation 2, becoming

$$Y[n,k] = f(X[n,k]) = \begin{cases} 0, & if\ |Im\{X[n,k]\}| \leq |Re\{X[n,k]\}| \\ |X[n,k]|^2, & if\ |Im\{X[n,k]\}| > |Re\{X[n,k]\}| \end{cases} \quad (2)$$

## VI. Test of the Input Activation Function

In order to test our proposed new approach to input activation functions, we chose a challenging real-world problem. There is increasing evidence that deep nets methods can analyze cough sounds of infected patients and predict COVID-19 [22]. Multiple research groups have gathered sound recordings for COVID-19 patients of all ages, in various settings, symptomatic and asymptomatic, and at different periods relative to symptom onset. These allow the trained deep net algorithms to learn audio characteristics of COVID-19 illness in patients with various demographic and medical conditions. A potential, purely digital COVID-19 detection method would allow for a smartphone-based rapid, equitable COVID-19 test with minimal infection risk, economic burden, and supply chain issues—all helpful factors to controlling COVID-19 spread. For details of our test system architecture see [23].

While working with various data sets for training, testing, and validation, we chose a challenging data set of recorded coughs collected by a large medical insurance company in Colombia by making phone calls to patients tested by PCR techniques for COVID-19. Most of these files were each under 4 seconds long. All data were zero-padded and truncated to an identical length of 4 seconds to set a fixed deep net input size. This data set, consisting of 7358 separate audio files, had a range of signal-to-noise (SNR) ratios of -10 to 28 dB. We chose to only keep samples with an estimated SNR higher than 15 dB, simulating a practical system which rejects unacceptably noisy inputs. After this selection we had 5284 remining files for the training, testing, and validation reported below.

We first trained and tested the deep net architecture and spectrogram parameter choices described in [23], with no change to the usual dB encoding of spectrograms, as labeled by $|X[n,k]|^2$ in equations 1 and 2. As shown in Figure 2 those results were compared to $Y[n,k]$ in equation 1. Moving to equation 2, with the cone boundary the same but the non-zero region rotated by $\phi = \frac{\pi}{2}$, resulted in very similar improvement. In fact, the symmetry is so strong that even though the retained points of $|X[n,k]|^2$ are completely different than those used to get the equation 1 result, the curves are very similar. Upon close inspection, the top blue curve is very close yet not identical to the top black dashed curve in Figure 2.

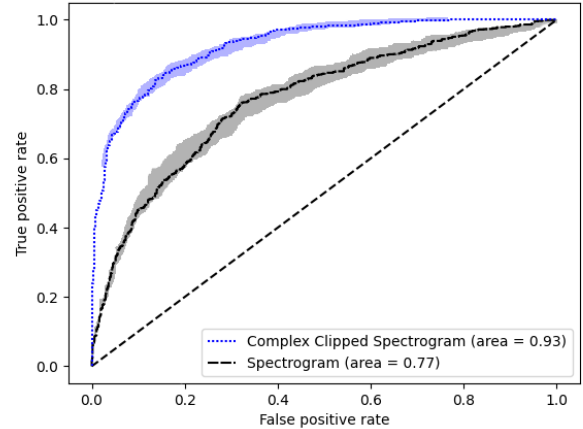

Fig. 2. Receiver operating curve for the baseline result, unprocessed $|X[n,k]|^2$ (solid black line) and our proposed complex clipping $Y[n,k] = f(X[n,k])$ (top blue dotted line) of equation 2 and (top dashed black line) of equation 3.) Confidence intervals are from the usual bootstrap method, e.g. [24][25].

For both the complex clipped spectrogram (equation 1) and the rotated version (equation 2), the area under the curve improved from 0.77 to 0.93 after complex clipping. This difference is potentially substantial for some applications. The performance curves of the complex clipped spectrogram and the rotated version (equation 2) were very close to each other yet were not identical.

We also found that empirical estimates of singular values of unprocessed versus complex clipped spectrograms showed that most estimated singular values, other than the largest singular value, were virtually unchanged by the non-linearity in equations 1 and 2. However the largest singular value was reduced by about 70%, suggesting that conditioning of the resulting $Y[n,k]$ representation was greatly improved, indicating that our proposed complex clipping effects a potentially useful regularization of the deep net at the input.

## VII. Conclusion

We have used the success of ReLU non-linearities between deep net layers to suggest a new form of ReLU, adapted to a complex to real mapping, from an STFT output to a modified spectrogram-like deep net input. We considered the desirable symmetries of this mapping and found that a cone-shaped support region could be used to remove deleterious redundancies from the input representation without degrading the input representation in other ways. The effectiveness of the removal of redundancies was confirmed via improved performance on an initial noisy acoustic data task.